\documentclass[12pt]{iopart}
\pdfoutput=1
\usepackage[numbers,sort&compress]{natbib}
\usepackage{amssymb}
\usepackage{graphicx}
\begin{document}
\title{Hubbard-corrected oxide formation enthalpies without adjustable parameters}
\author{J Voss}
\address{SUNCAT Center for Interface Science and Catalysis, SLAC National Accelerator Laboratory\\ Menlo Park, CA 94025, USA}
\ead{vossj@slac.stanford.edu}

\begin{abstract}
A density functional theory (DFT) approach to computing transition metal oxide heat of formation without adjustable parameters is presented. Different degrees of $d$-electron localization in oxides are treated within the DFT+$U$ approach with site-dependent, first-principles Hubbard $U$-parameters obtained from linear response theory, and delocalized states in the metallic phases are treated without Hubbard corrections. Comparison of relative stabilities of these differently treated phases is enabled by a local $d$-electron density matrix-dependent model, which was found by genetic programming against experimental reference formation enthalpies. This mathematically simple model does not explicitly depend on the Hubbard-corrected ionic species and is shown to reproduce the heats of formation of the Mott insulators Ca$_2$RuO$_4$ and Y$_2$Ru$_2$O$_7$ within $\sim$3\% of experimental results, where the experimental training data did not contain Ru oxides. This newly developed method thus absolves from the need for element-specific corrections fitted to experiments in existing Hubbard-corrected approaches to the prediction of reaction energies of transition metal oxides and metals. The absence of fitting parameters opens up here the possibility to predict relative thermodynamic stabilities and reaction energies involving $d$-states of varying degree of localization at transition metal oxide interfaces and defects, where site-dependent $U$-parameters will be particularly important and devising a fitting scheme against experimental data with predictive power would be exceedingly difficult.
\end{abstract}
\noindent{\it Keywords\/}: {density functional theory, machine learning, strongly correlated systems, transition metal oxides}\\
\maketitle

\section{Introduction}

The Kohn-Sham formulation \cite{KohnPR1965} of density functional theory (DFT) \cite{HohenbergPR1964} is widely used to model the electronic structures of bulk solids, surfaces, and molecules \cite{BurkeJCP2012,BeckeJCP2014,JonesRMP2015}. With a reasonable trade-off between accuracy and computational complexity, the possibility to treat the electronic structures of solids as Bloch states of the Kohn-Sham single-particle-like Hamiltonian \cite{PayneRMP1992} renders DFT and extensions the de-facto standard for electronic structure simulation in computational materials science \cite{CurtaroloCMS2012,SaalJOM2013,JainAPLM2013,GhiringhelliNCM2017,TalirzSD2020}.

\begin{figure}
\includegraphics[width=\textwidth,keepaspectratio=true]{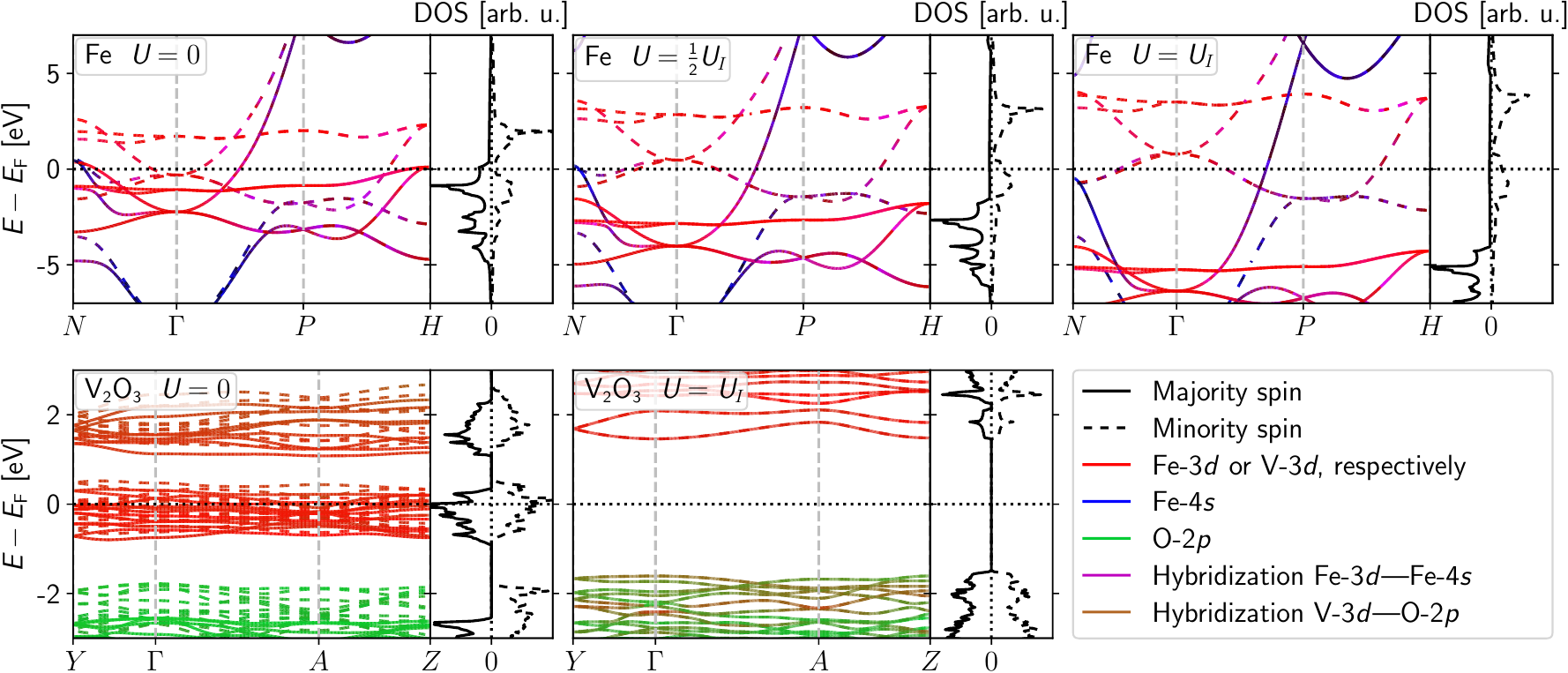}
\caption{Bandstructures and densities of states of Fe (top) and V$_2$O$_3$ (bottom) computed using GGA and GGA+$U$. $U_I$ is the system-specific Hubbard parameter obtained from linear response ($U_I^{\rm Fe}{\approx}7.2$~eV and $U_I^{\rm V}{\approx}6.0$~eV; details in main text). Already at half this parameter strength, Fe shows an incorrect suppression of majority spin density of states at the Fermi level $E_{\rm F}$ and tendency towards spin saturation (the magnetic moment is increased by about 30\% compared to the $U=0$ case). At full $U_I$ strength, the magnetic moment increase is 40\%. For V$_2$O$_3$, on the other hand, GGA incorrectly predicts a ferromagnetic metallic phase as the ground state (bandstructure plot folded into the smaller irreducible Brillouin zone of the antiferromagnetic phase), while GGA+$U$ correctly predicts an antiferromagnetic Mott-insulating phase.\label{fig:bandstructures}}
\end{figure}

Generalized gradient approximations (GGA) and meta-GGAs to electronic exchange and correlation (XC) in Kohn-Sham DFT only depend locally on the charge density, its gradient and Laplacian, and the Kohn-Sham kinetic energy density and yield reasonable cohesive energies with typical errors in comparison to experiment of $\lesssim$0.2~eV per atom for those solids \cite{ZhangNJP2018,BrownJCC2021} where self-interaction errors due to these semi-local approximations are not too large. In materials with strong electron localization, these errors become large, and semi-local DFT thus typically yields significantly larger per atom errors for relative stabilities of {\it e.g.}\ transition metal (TM) oxide phases \cite{WangPRB2006}. The failure of DFT also manifests itself in a qualitatively wrong description of the electronic structure of Mott insulators, where the semi-local, mean-field treatment of XC incorrectly yields a metallic ground state (see Fig.~\ref{fig:bandstructures}).

One approach to reducing self-interaction errors is to admix a fraction of Hartree-Fock exchange energies computed from the Kohn-Sham orbitals to the semi-local XC functional \cite{BeckeJCP1993}. The computational cost for evaluating the non-local Fock-operator generally increases the cost for this hybrid DFT method by a factor of the system size in comparison to semi-local DFT. In contrast to unscreened Hartree-Fock, which unphysically predicts diverging Fermi velocity and vanishing density of states at the Fermi level for the homogeneous electron gas \cite{Fulde1995}, screened hybrid DFT \cite{HeydJCP2003} can treat metallic phases. Magnetic metals are however described worse, {\it e.g.}, with the widely used HSE06 screened hybrid functional \cite{HeydJCP2006} than with semi-local DFT, with overestimated magnetic moments of Fe, Co, and Ni \cite{JanthonJCTC2014,GaoSSC2016}, incorrectly predicting bcc and hcp ground state structures for Co and Ni \cite{JangJPSJ2012}, respectively, and describing surface reaction energetics worse in particular for magnetic TM surfaces \cite{SharadaPRB2019}. HSE06 simulations were nevertheless shown to improve on transition metal oxide formation energies over GGA-DFT \cite{ChevrierPRB2010}, showing the importance of the reduction of self-interaction errors due to localized electrons in the oxides.

Another, computationally more efficient approach to addressing the self-interaction errors is the GGA+$U$ approach \cite{AnisimovPRB1991,CzyzykPRB1994,LiechtensteinPRB1995}, which does not lead to an overall increase in computational complexity over semi-local DFT. In the GGA+$U$ method, the Kohn-Sham Hamiltonian is supplemented with a Hubbard-like term, which depends on the local projections of the Kohn-Sham single-particle density matrix at ionic sites with strong electron localization. The Hubbard term depends thus explicitly on the Kohn-Sham occupation numbers and can {\it e.g.}\ open up a particle-hole excitation gap in the case of Mott insulators. The strength of the Hubbard term is given by the parameter $U$, which can either be considered as an adjustable, to be fitted parameter or be computed from first principles \cite{GunnarssonPRB1989,CococcioniPRB2005,MiyakePRB2008}. In recent approaches, machine learning techniques were employed to find $U$-parameters to mimic hybrid DFT bandstructures and bandgaps with GGA+$U$ \cite{YuNCM2020} or to improve agreement with experimental lattice constants, magnetic moments, and bandgaps \cite{TavadzeNCM2021}.

A drawback of the GGA+$U$ method is that the ground state energies of Hamiltonians with different Hubbard parameters $U$ at a given ionic site cannot be compared to each other. The compromise of choosing average $U$-parameters for each transition metal ionic species (irrespective of {\it e.g.}\ differences in oxidation state) leads to an improvement of oxidation energies of transition metal oxides over GGA-DFT \cite{WangPRB2006,GautamPRM2018,LongPRM2020}. For molecules, meaningful total energy differences in the case of first-principles, reactant-dependent $U$-values can be computed by integrating the forces acting on the ions along a reaction path \cite{KulikJCP2011}. For the prediction of oxidation energies of bulk TM oxides with first-principles $U$-parameters, the unit cell volume-dependence of the $U$-parameters can be computed, and integrals over the volume changes required to match the $U$-values of different oxides can then be used to estimate reaction energies with the oxides at their respective equilibrium volumes \cite{XuJCP2015}. Oxidation energies were found to be further improved using this method rather than using average $U$-parameters, while oxide formation enthalpies could not be improved \cite{XuJCP2015}. A general problem for either an average $U$-parameter approach or relying on volume change integrals is that first-principles methods to computing $U$ yield nonzero values also for metallic phases. Delocalized metallic states are however better described without supplemental Hubbard terms (see Fig.~\ref{fig:bandstructures} for an example of the unphysical effects due to application of a $U$-term to a metallic bandstructure).

Improvement over TM oxide formation energy predictions from screened hybrid DFT was achieved by treating the oxide phases at the GGA+$U$ level with average $U$-parameters for the TM species and the metallic phases within the GGA \cite{JainPRB2011}. For each Hubbard-corrected TM species, an energy offset was fitted to experimental formation enthalpies, so that GGA+$U$ total energies plus offset can be compared to the GGA total energies for the metallic phases. These fitted offsets allow the resulting corrected total energy differences between GGA+$U$ and GGA calculations to be interpreted as reaction energies. In a similar approach, elemental reference energies were fitted directly in Ref.~\cite{StevanovicPRB2012} against experimental formation enthalpies with TM oxides treated at the GGA+$U$ level with average $U$-parameters, and formation energies predicted with this approach were used to fit a multivariate linear regression model of formation energies exclusively against atomic properties \cite{DemlPRB2016}. With the availability of sufficient experimental reference data, the approach of combining GGA and GGA+$U$ total energies can be further refined by fitting offset energies not only for different TM species, but also for different oxidation states and different ligand coordination \cite{AykolPRB2014}. While experimental calorimetry reference data for thermodynamics of stoichiometric bulk TM oxides can be used here to parametrize these element- or even system-specific corrections, determining the required offset energies for Hubbard-corrected TM sites near {\it e.g.}\ point defects or interfaces will be difficult.

Therefore, a newly developed method free of adjustable parameters is presented here that combines the treatment of $d$-electrons of varying degrees of localization within the GGA+$U$ approach with site-specific, first-principles $U$-parameters from linear response for TM oxides with GGA calculations optimal for metallic phases. To correct for the offset between GGA+$U$ total energies with different $U$-parameters and GGA ({\it i.e.}\ $U$=0) total energies, genetic programming with experimental TM oxide formation energies as references is applied to find a mathematical model that does not explicitly depend on the Hubbard-corrected TM species. The resulting, relatively simple mathematical model does thus not have adjustable, system-specific parameters. Examples of the applicability of the model for TM oxides with Hubbard-corrected TM species not contained in the training data are shown.

\section{Method}

\begin{figure}
\centering
\includegraphics[width=3.30in,keepaspectratio=true]{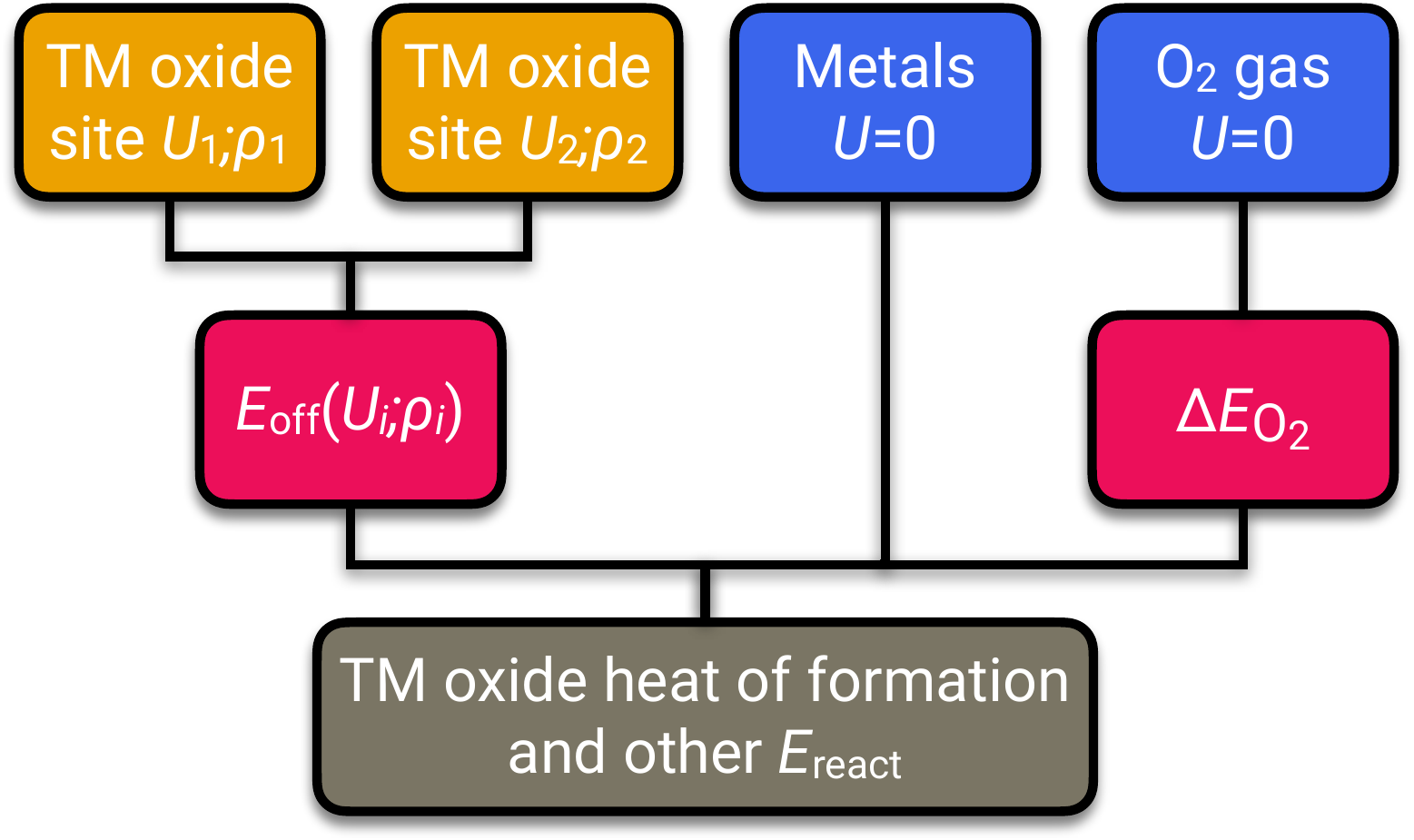}
\caption{Schematic of the presented method to computing TM oxide heat of formation without adjustable parameters. TM oxide phase total energies are computed using GGA+$U$ with site-dependent, first-principles $U$-parameters, and metallic phases and the O$_2$ gas phase reference are treated without Hubbard corrections. The O$_2$ total energy reference is corrected adding a constant offset of $\sim$0.84~eV as described in the main text. Based on energy offsets from a genetic programming model, the total energies from GGA+$U$ with site-dependent $U$-parameters are adjusted, so that comparison to the GGA total energies of the metallic phases becomes possible and reaction energies involving both oxides and metals can be computed.\label{fig:schematic}}
\end{figure}

The method consists of computing DFT and DFT+$U$ total energies for metal and oxide phases, respectively, as described below and using a genetic programming model for finding an energy offset for the Hubbard-corrected oxide total energies, such that physically meaningful energy differences between metals and oxides can be obtained (Fig.~\ref{fig:schematic}).

\subsection{DFT calculations}

GGA and GGA+$U$ simulations are performed with the VASP code \cite{KressePRB1996} using plane-wave basis sets with a cut-off energy of 600~eV and representing ionic cores by projector-augmented wave (PAW) \cite{BloechlPRB1994} frozen-core pseudopotentials \cite{KressePRB1999}. Brillouin zones are sampled with $\Gamma$-point centered, regular grids with a spacing of at most 0.02~\AA$^{-1}$, and Gaussian smearing with a width of 0.05~eV is used to determine Kohn-Sham occupation numbers. The widely used PBE functional \cite{PerdewPRL1996} is employed for the GGA to electronic exchange and correlation. Forces acting on ions and stress on the unit cells are simultaneously relaxed, with residual forces of less than 0.02~eV/{\AA} per atom and residual pressures of less than 0.02~GPa, respectively.

The total energy of the O$_2$ molecule is described poorly by GGA approaches, which is typical for multiply bonded molecules \cite{GunnarssonPRB1989,ErnzerhofIJQC1997}. Empirical approaches fit the total energy of O$_2$ to experimental reaction energies, such as oxidation energies of solids \cite{WangPRB2006} or the formation of H$_2$O \cite{NorskovJPCB2004}. Here, instead of empirically fitting to the experimental formation energy of H$_2$O, atomization energies as computed with the first-principles Weizmann-4 computational thermochemistry protocol (W4) \cite{KartonJCC2017} are referenced. Using the difference of the atomization energies of O$_2$ and H$_2$O
\begin{equation}
E^{\rm atom}({\rm O}_2) - 2E^{\rm atom}({\rm H}_2{\rm O}) = E({\rm O}_2) - 2E({\rm H}_2{\rm O}) + 4E({\rm H}) ,
\end{equation}
the total energy of O$_2$ used in this study is computed as
\begin{equation}
E({\rm O}_2) = E^{\rm atom}_{\rm W4}({\rm O}_2) - 2E^{\rm atom}_{\rm W4}({\rm H}_2{\rm O}) + 2E_{\rm GGA}({\rm H}_2{\rm O}) - 4E_{\rm GGA}({\rm H}) \label{eq:eo2}.
\end{equation}
$E^{\rm atom}({\rm O}_2)$ and $E^{\rm atom}({\rm H}_2{\rm O})$ are the atomization energies of the molecules ${\rm O}_2$ and ${\rm H}_2{\rm O}$, respectively, and the subscript ${\rm W4}$ indicates calculation of the atomization energies at the W4 level of theory. $E({\rm O}_2)$, $E({\rm H}_2{\rm O})$, and $E({\rm H})$ are the total energies of ${\rm O}_2$, ${\rm H}_2{\rm O}$, and H, respectively, and the subscript GGA indicates calculation of these energies at the GGA level of theory.
The GGA total energies $E_{\rm GGA}({\rm H})$ and $E_{\rm GGA}({\rm H}_2{\rm O})$ are obtained with supercells separating the periodic images by $\sim$16~{\AA} (without Brillouin zone sampling or occupation broadening). The total energy computed with (\ref{eq:eo2}) is about $0.84$~eV higher ({\it i.e.}~less stable) than $E_{\rm GGA}({\rm O}_2)$.

\subsection{Hubbard corrections}

\subsubsection{DFT+$U$ energy functional~~}

In the DFT+$U$ approach, the DFT energy functional $E_{\rm DFT}[n(\textbf{r})]$ (considered here to employ a GGA) of the electron charge density $n(\textbf{r})$ is supplemented with a Hubbard-like term:
\begin{equation}
E_{{\rm DFT}+U}[n(\textbf{r}),\rho_I] = E_{\rm DFT}[n(\textbf{r})] + E_{\rm Hub}[\rho_I] - E_{\rm DC}[\rho_I] \label{eq:dftpu}.
\end{equation}
$\rho_I$ are projections of the Kohn-Sham density matrix onto atomic-like ({\it i.e.}\/, $3d$, $4d$ or $5d$) orbitals centered on the transition metal ions $I$. $E_{\rm DC}[\rho_I]$ is a term only depending on the total projected spin occupations at each site $I$ and corrects for partially double-counting Coulomb interactions in $E_{\rm DFT}[n(\textbf{r})]$ and the Hubbard-term $E_{\rm Hub}[\rho_I]$. Here, the approach of Anisimov {\it et al.}~\cite{AnisimovPRB1991} and Liechtenstein {\it et al.}~\cite{LiechtensteinPRB1995} is followed: $E_{\rm DC}[\rho_I]$ is chosen such that it cancels $E_{\rm Hub}[\rho_I]$ in case of integer eigenvalues of the projected density matrices $\rho_I$, {\it i.e.}\/, in the limit of isolated ions. For $E_{\rm Hub}[\rho_I]$, the spherically averaged expression introduced by Dudarev {\it et al.}~\cite{DudarevPRB1998} is chosen:
\begin{equation}
E_{\rm Hub}[\rho_I] = \sum_I \left( \frac{\bar U_I}{2}N_I^2 - \frac{\bar J_I}{2} \!\sum_{\sigma\in\{\uparrow,\downarrow\}}\! {\left(N_I^\sigma\right)}^2 - \frac{\bar U_I - \bar J_I}{2} \!\sum_{\sigma\in\{\uparrow,\downarrow\}}\! {\rm Tr}\, {\left( \rho_I^\sigma \right)}^2 \right) \label{eq:ehub},
\end{equation}
and
\begin{equation}
E_{\rm DC}[\rho_I] = \sum_I \left( \frac{\bar U_I}{2} N_I(N_I-1) - \frac{\bar J_I}{2} \!\sum_{\sigma\in\{\uparrow,\downarrow\}}\! N^\sigma_I(N^\sigma_I-1) \right) ,
\end{equation}
such that
\begin{equation}
E_U[\rho_I] = E_{\rm Hub}[\rho_I] - E_{\rm DC}[\rho_I] = \sum_I \frac{\bar U_I - \bar J_I}{2} \left( \sum_{\sigma\in\{\uparrow,\downarrow\}}\! {\rm Tr}\, \rho_I^\sigma - {\rm Tr}\, {\left( \rho_I^\sigma \right)}^2 \right) \label{eq:e_u}.
\end{equation}
${\rm Tr}\, \rho^\sigma = \sum_{m=-\ell}^\ell \rho_{mm}^\sigma$ is the trace over the orbital degrees of freedom of the projected density matrices ($2\ell+1=5$ in the case of $d$-orbitals). $\sigma$ denotes the electron spin degree of freedom (spins assumed to be collinear in this study), and $N_I = \sum_\sigma N_I^\sigma = \sum_\sigma {\rm Tr}\, \rho_I^\sigma$ is the total projected occupation at site $I$. $\bar U_I$ and $\bar J_I$ are the spherically averaged screened on-site Coulomb repulsion and exchange interactions at site $I$, respectively. Since Eq.~\ref{eq:e_u} depends on the difference between these interactions at each site $I$, it is convenient to introduce an effective Hubbard parameter $U_I$ with
\begin{equation}
U_I = \bar U_I - \bar J_I \label{eq:u_i}.
\end{equation}
The Kohn-Sham system is exposed to the following additional potential due to the Hubbard correction at site $I$:
\begin{equation}
V^{I,\sigma}_{mm'} = \frac{\partial E_U}{\partial \rho^{I,\sigma}_{m'm}} = U_I \left( \frac{1}{2} \delta_{mm'} - \rho^{I,\sigma}_{mm'}\right) \label{eq:v_u}.
\end{equation}
The Hubbard corrections (\ref{eq:e_u}) effectively constitute an energy penalty for non-integer eigenvalues of $\rho_{I,\sigma}$. Considering Eq.~\ref{eq:v_u} in the eigenbasis of $\rho^{I,\sigma}$, the potential is positive for projected occupation eigenvalues less than $1/2$, favoring a lowering of the eigenvalue towards zero. For eigenvalues larger than $1/2$, the potential is negative thus favoring increasing the eigenvalue towards one. Accordingly, the Hubbard corrections with their explicit occupation number dependence can open a Mott-Hubbard gap in the Kohn-Sham spectrum of an otherwise metallic system in a single-particle bandstructure picture.

\subsubsection{Hubbard-$U$ from first principles~~}

Rather than treating the effective Hubbard parameters $U_I$ as empirical fitting parameters, the linear response approach to determining $U_I$ by Cococcioni and de Gironcoli \cite{CococcioniPRB2005} was implemented here on top of the existing DFT+$U$ code in VASP. DFT total energies spuriously show a non-linear dependence on fractional electron numbers (fractional particle numbers correspond to quantum mechanical ensemble averages, and the total energies should show linear behavior between pairs of adjacent integer particle numbers). The $U_I$ can be chosen to compensate spurious curvature with respect to changes in the Hubbard site occupations $N_I$. Instead of constraining Hubbard site occupations to determine these curvatures (which is difficult with a plane-wave basis set), the linear response approach adds perturbations to the potentials (\ref{eq:v_u}): $\tilde V^{I,\sigma}_{mm'} = V^{I,\sigma}_{mm'} + \alpha_I \delta_{m,m'}$. The functional (\ref{eq:dftpu}) is Legendre transformed such that the energy is minimized with respect to the potential perturbations $\alpha_I$ instead of the Hubbard-site occupation numbers, and the Hubbard parameters $U_I$ are then determined as \cite{CococcioniPRB2005}:
\begin{equation}
U_I = \left( \chi_0^{-1} - \chi^{-1} \right)_{II} \label{eq:uresp}.
\end{equation}
$\chi_{IJ} = \partial N_I/\partial\alpha_J$ is a density response function of the total occupations $N_I$ of each Hubbard site $I$ to perturbations from the $\alpha_J$ for any of the Hubbard sites $J$ and is computed here with finite differences in $\alpha_J$, letting the electronic structure self-consistently adjust to the perturbed Hubbard potentials. Not letting the Kohn-Sham potentials self-consistently adjust to the perturbed Hubbard potentials, but only computing the Kohn-Sham bands and thus the Kohn-Sham density matrix at the fixed perturbed potentials leads to the perturbed Hubbard site occupations $N_I^0$. The finite difference $\chi^0_{IJ} = {\Delta}N_I^0/\Delta\alpha_J$ is used to compute the corresponding bare response function.

While the updated Kohn-Sham potentials can self-consistently screen the effect of the perturbed Hubbard potentials, the bare response accounts for single-particle bandstructure-only rehybridization between the Hubbard sites and their surroundings. The resulting curvature in total energy due to these latter rehybridizations is not a measure of a self-interaction error to be corrected for with a Hubbard-term, and Eq.~\ref{eq:uresp} thus removes this curvature from the computation of the $U_I$.

\subsubsection{\label{sec:upractice}{\it Ab initio}~$U$-parameters in practice~~}

In the present implementation, the unperturbed electronic structure is computed first, then the perturbation due to $\alpha_J$ is added, and the Kohn-Sham bands are updated yielding the bare response $\chi^0_{IJ}$ for all sites $I$, and finally this same simulation is continued allowing the Kohn-Sham potentials to adjust yielding the response $\chi_{IJ}$ for all sites $I$. The Spglib code \cite{TogoARXIV2018} is used to identify symmetrically equivalent Hubbard sites and the corresponding symmetry operations, so that simulations only need to be performed for perturbations at nonequivalent sites and the full $\chi$ and $\chi^0$ response matrices can be obtained using symmetry operations.

To decouple the effect of the perturbation from its periodic images, the TM oxide unit cells are repeated here into supercells separating the periodic site images by at least $\sim$8~{\AA}. Coccocioni and de Gironcoli \cite{CococcioniPRB2005} find that extending the Hubbard sites by a fictitious background site for charge neutrality of the perturbations improves convergence with respect to supercell size. The response function matrices $\chi$ and $\chi^0$ are thus extended by one row and one column, and the additional matrix elements are determined by the charge neutrality requirement of the sums over each row and each column of the matrices yielding zero. With this construction, the extended response matrices now have a zero singular value and cannot be inverted. This zero singular value is due to the fact that adding a constant offset to all Hubbard site potentials and the background does not lead to any total energy curvature. Here, the Moore-Penrose pseudo inverse is used instead of the matrix inverses in Eq.~\ref{eq:uresp}, such that this zero singular value leads to a zero singular value for the pseudo inverse, correctly not affecting the computed $U_I$.

After the $U_I$ have been computed in supercells, GGA+$U$ optimizations of the primitive cells with respect to atomic forces and unit cell stress are performed. Based on the optimized structures new supercells are created to re-compute the $U_I$ and then to re-optimize the primitive cells again. Performing this relaxation and $U_I$ determination cycle twice led to mean average TM oxide lattice constant changes $< 10^{-2}$~{\AA}, and thus no further structural-dependence self-consistency cycles of the $U_I$ have been performed.

Generally, the $U_I$ are computed here from GGA$+U$ calculations with all Hubbard parameters set to zero, yielding the spurious energy curvatures to correct for. Some of the considered TM oxides are incorrectly predicted as metallic at the GGA level. For these oxides, GGA$+U$ simulations are performed at a range of Hubbard parameters $U^{\rm input}$ large enough to open a bandgap, and, following the idea of Ref.~\cite{KulikPRL2006}, the computed $U_I$ values are linearly extrapolated to $U^{\rm input}=0$ (see Table S1 for a list of all computed $U_I$ values, indicating which TM oxide cases are based on this extrapolation technique).

The values of the Hubbard-parameters $U_I$ determined {\it via}\ Eq.~\ref{eq:uresp} have a dependence on the choice of the projectors of the Kohn-Sham density matrix on the Hubbard sites. The VASP code uses the projections of the Kohn-Sham pseudo wavefunctions onto the PAW projector functions, which are non-zero only inside the augmentation spheres around the ions \cite{RohrbachJPCM2003,BengonePRB2000}. The DFT+$U$ implementation in the Quantum Espresso plane-wave pseudopotential DFT code, {\it e.g.}\/, offers further choices for the definition of the projection onto the Hubbard sites, such as the $d$-orbitals of free atoms with or without orthogonalization with respect to the atomic orbital overlap between Hubbard sites \cite{GiannozziJPCM2009}. Using the density functional perturbation implementation \cite{TimrovPRB2021} in the Quantum Espresso suite to computing $\chi$ and $\chi^0$ with ultrasoft pseudopotentials \cite{VanderbiltPRB1990} from the GBRV dataset \cite{GarrityCMS2014}, the computed $U$ for bulk Fe is about 7.0~eV and 5.5~eV with and without projector orthogonalization, respectively. Similarly, $U$-parameters obtained from constraining local occupations were found to depend strongly on TM ion muffin-tin sphere radii \cite{NawaPRB2018}. These dependencies show the importance of determining the parameters $U_I$ with the same projectors as are being used in the DFT+$U$ simulation.

Similarly large $U$-parameters for bulk Fe based on constrained DFT are reported in Ref.~\cite{AnisimovPRB1991b}, and also the VASP PAW implementation (without inter-site projector overlap due to non-overlapping PAW augmentation spheres) finds a large value of about 7.2~eV. Even only applying half this Hubbard-parameter strength in a GGA+$U$ simulation of bulk Fe leads to unphysical suppression of majority spin density of states around the Fermi level (see Fig.~\ref{fig:bandstructures}) and the magnetization is spuriously driven towards saturation. As the corresponding $d$-band shifts would deteriorate the description of, {\it e.g.}\/, transition metal surface chemistry \cite{SharadaPRB2019}, this example highlights that GGA+$U$ simulations of metallic systems with first-principles $U$-parameters should be avoided because of potentially unphysical predictions.

A promising direction for (at least partial) compensation of unphysical effects on metallic electronic structures in DFT+$U$ is the DFT+$U$+$V$ approach \cite{LeiriaCampoJPCM2010}, where also inter-site Hubbard corrections, determined from the off-diagonal elements in (\ref{eq:uresp}), are applied. The DFT+$U$+$V$ method is implemented in Quantum Espresso, and for the above orthogonalized projector case with an on-site Hubbard parameter of 7.0~eV, the $V$-parameter between nearest neighboring sites is only $\sim$0.3~eV, barely affecting the electronic structure qualitatively and both GGA+$U$ and GGA+$U$+$V$ calculations yielding a spuriously saturated, integer magnetization of $\sim$3 Bohr magnetons per Fe atom. Generally, the DFT+$U$+$V$ method can be extended to such off-diagonal Hubbard terms between $d$- and $p$-projections, also on the same site \cite{LeiriaCampoJPCM2010}, which could improve suppression of unphysical effects of Hubbard-$U$ on metallic systems, but here only on-site Hubbard-corrections based on $d$-projections are considered and implemented. All results presented in the following treat metallic phases without Hubbard corrections.

\subsection{\label{sec:refdata}Reference data}

\subsubsection{TM oxide and metallic phases~~}

To find a model enabling the energetic comparison of products and reactants with first-principles $U$-parameters rather than average $U$-values, experimental formation enthalpies of transition metal oxides requiring Hubbard corrections are used as training data with corrections as described below. The formation enthalpies of 66 such TM oxides containing the TM ions V, Cr, Mn, Fe, Co, Ni, and Mo are taken from Ref.~\cite{Kubaschewski1993}. A Debye model fitting technique introduced and benchmarked in Ref.~\cite{HautierPRB2012} is used to determine $T=0$~K enthalpies from the experimental values at 298~K. Neglecting thermal expansion of the solid phases, the heat capacity at constant pressure
\begin{equation}
c_p(T) = y \cdot{\left(\frac{T}{T_{\rm D}}\right)}^3 \int\limits_0^{T/T_{\rm D}} \frac{u^4 \exp(u)}{{\{1-\exp(u)\}}^2} {\rm d}u \label{eq:cp}
\end{equation}
and the entropy
\begin{equation}
S(T) = \int\limits_0^T \frac{c_p(T')}{T'} {\rm d}T' \label{eq:sint}
\end{equation}
are fitted numerically to match the $T=298$~K entropies and heat capacities provided in Ref.~\cite{Kubaschewski1993} ($c_p(298\,{\rm K})$ and $S(298\,{\rm K})$ for ZnFe$_2$O$_4$ taken from Ref.~\cite{WestrumJPCS1957} due to unphysical $c_p(298\,{\rm K})$ in \cite{Kubaschewski1993}), both for the TM oxides and the metallic reference phases, considering the Debye temperature $T_{\rm D}$ and the pre-factor $y$ as fitting parameters.

Here, the fitted $T_{\rm D}$ is furthermore employed to provide an experimental estimate of the phonon zero-point energy $E_{\rm ZP}$ per formula unit using that the Debye model density of states is quadratic in the phonon frequency $\omega$ below the Debye cut-off frequency $\omega_{\rm D}=k_{\rm B}T_{
\rm D}/\hbar$:
\begin{equation}
E_{\rm ZP} = \frac{\hbar}{2} \frac{9N_{\rm A}}{\omega_{\rm D}^3} \int\limits_0^{\omega_{\rm D}} \omega^3 {\rm d}\omega = \frac{9}{8} N_{\rm A} k_{\rm B} T_{\rm D}.
\end{equation}
$N_{\rm A}$ is the number of atoms per formula unit, $\hbar$ the reduced Planck constant, and $k_{\rm B}$ Boltzmann's constant. The respective enthalpy changes $\int_0^{298\,{\rm K}}c_p(T){\rm d}T$ and the relatively small $E_{\rm ZP}$ are subtracted from the formation enthalpies for each of the solid product and reactant phases (for the metallic reactant phases considered here, the enthalpy of formation at 298~K is zero by definition). 

\subsubsection{O$_2$ reference~~}

From the zero enthalpy reference of O$_2$ at standard conditions the integral $\int_{90\,{\rm K}}^{298\,{\rm K}}c_p(T){\rm d}T$ is subtracted (heat capacity data valid in this temperature range taken from Ref.~\cite{NIST}). For the remaining integral from 0~K to 90~K to be subtracted, a constant $c_p = 5R/2$ ($R$: molar gas constant) is assumed. Furthermore, the zero-point energy of O$_2$ as computed with the W4 quantum chemistry protocol is subtracted \cite{KartonJCC2017}.

Assuming some degree of cancellation of pressure times volume ($pV$) terms between solid reactants and products \cite{HautierPRB2012}, $pV$ terms are neglected for the solid phases. The formation energy training targets are thus given as the experimental enthalpy differences between the TM oxide and metallic phases and O$_2$ with heat capacity integrals and zero-point energies subtracted for all phases, as described above.

\subsubsection{\label{sec:ruoxbench}Ru oxide benchmarks~~}

As test cases outside the training and validation data, two correlated Ru oxides with available experimental thermodynamic data are considered: Ca$_2$RuO$_4$ and Y$_2$Ru$_2$O$_7$. The experimental enthalpy of Ca$_2$RuO$_4$ at 1140~K relative to $2\,$CaO and RuO$_2$ is taken from Ref.~\cite{JacobJES2003}. The formation enthalpies of CaO and RuO$_2$ and their entropies and heat capacity data for Debye model fits and $c_p(T)$ integration are taken from Ref.~\cite{Kubaschewski1993}, so that O$_2$ and solid Ca and Ru can be used as reference phases (with zero-point energies from Debye model fits to data in \cite{Kubaschewski1993} subtracted from the Ca and Ru references and the same adjustments for O$_2$ as described above). The Debye temperature of Ca$_2$RuO$_4$ is determined from the curvature in low-temperature $c_p(T)/T$ measurements in Ref.~\cite{NakatsujiPHD2000} to be $T_{\rm D}{\approx}418$~K. The entropy of Ca$_2$RuO$_4$ is computed as $S(298\,{\rm K})=\int_0^{298\,{\rm K}} c_p(T)/T{\rm d}T$, where $c_p(T)$ is taken from Ref.~\cite{QiPRB2012} for Ca$_2$Ru$_{1-x}M_x$O$_4$ with $M\in\{\textrm{Mn},\textrm{Fe}\}$. The estimated entropy at 298~K is about 147~JK$^{-1}$mol$^{-1}$, irrespective of different small values of $x\lesssim1\%$. Given the estimated $T_D$, the parameter $y$ is fitted to yield the entropy at 298~K. In analogy to the other solid phases, the integrated heat capacity and the Debye-model zero-point energy are subtracted from the enthalpy of Ca$_2$RuO$_4$.

The experimental enthalpy of formation of Y$_2$Ru$_2$O$_7$ at 298~K is taken from Ref.~\cite{BanerjeeJSSE2019}. Debye model fits for the corrections of Y and Ru are based on data in \cite{Kubaschewski1993}, and also O$_2$ is corrected analogously to the other oxide heats of formation. A low-temperature heat capacity fit yielding a Debye temperature of 449~K for Y$_2$Ru$_2$O$_7$ is taken from Ref.~\cite{BlacklockJCP1980}. Low and moderate-temperature heat capacity data from Refs.~\cite{TairaJSSC2000} and \cite{BanerjeeJSSE2019}, respectively, is used in the integral (\ref{eq:sint}) to determine the entropy at 298~K of $\sim$216~JK$^{-1}$mol$^{-1}$ and fit the parameter $y$ of the Debye model for $T=0$ extrapolation and subtraction of $E_{\rm ZP}$. The resulting benchmark Ru oxide formation energies and the $3d$-TM and Mo oxide formation energies used for the model search are listed in Table S2 in the SI.

\subsection{Model search}

The {\it ansatz}~for oxide formation energies in the model search used here is based on an energy offset $E_{\rm off}(U_I,\rho_I)$ for every correlated site $I$:
\begin{eqnarray}
E^{\rm form} & = & E_{{\rm GGA}+U}({\rm TM~oxide}) - \sum_I E_{\rm off}(U_I,\rho_I) \nonumber\\
& -\! & \!\sum_{\nu=1}^{N_{\rm metals}} \beta_\nu E_{\rm GGA}({\rm metal~number}~\nu) - \frac{\gamma}{2} E({\rm O}_2) \label{eq:eform},
\end{eqnarray}
where the sum over $\nu$ accounts for the $N_{\rm metals}$ metallic references calculated at the GGA level of theory and the factor $\beta_\nu$ for the stoichiometric occurrence of these metals in the oxide. $\gamma$ accounts for the stoichiometry of O in the oxide, and $E({\rm O}_2)$ is the corrected value from Eq.~\ref{eq:eo2}. The offset at site $I$ is chosen to be of the form
\begin{equation}
E_{\rm off}(U_I,\rho_I) = \frac{U_I}{2} f_{\rm GP}\!\left(\sum_\sigma{\rm Tr}\rho^\sigma_I,\sum_\sigma{\rm Tr}{\left(\rho^\sigma_I\right)}^2\right) \label{eq:eoff}.
\end{equation}
$f_{\rm GP}$ is a function of the sum of the eigenvalues of the projected density matrix for site $I$ and of the sum of the squares of these eigenvalues, ensuring rotational invariance of the model. Sums of higher eigenvalue powers as additional arguments for $f_{\rm GP}$ did not lead to improved models, and models depending on higher eigenvalue powers hence are not presented in the following.

As $f_{\rm GP}$ has no explicit site-dependence, site indices $I$ are dropped in the following from the density matrices and Hubbard parameters for simplicity, noting that evaluating a model always corresponds to a sum over all correlated sites $I$ in the oxide. Starting with the initial guess
\begin{equation}
f^0_{\rm GP}\!\left(\sum_\sigma{\rm Tr}\rho^\sigma,\sum_\sigma{\rm Tr}{\left(\rho^\sigma\right)}^2\right) = \sum_\sigma{\rm Tr}\rho^\sigma - \sum_\sigma{\rm Tr}{\left(\rho^\sigma\right)}^2 \label{eq:start},
\end{equation}
corresponding to $E_{\rm off}(U,\rho)$ equaling the density matrix-dependent part $E_U(U,\rho)$ (Eq.~\ref{eq:e_u}) of the DFT+$U$ functional (\ref{eq:dftpu}), genetic programming \cite{Koza1992,KozaSC1994} is used to evolve generations of programs $\{f^z_{\rm GP}\}$ with the gplearn code \cite{StephensGP2019}.

As a measure of fitness for ranking the models $f^z_{\rm GP}$, the Akaike information criterion (AIC) \cite{AkaikeISIT1973} is implemented:
\begin{equation}
{\rm AIC} = -2 \mathfrak{L} + 2 \Phi \label{eq:aic},
\end{equation}
where $\mathfrak{L}$ is the maximum of the log-likelihood function for the considered model and $\Phi$ is the number of free parameters. Under the assumption of $n$ normally and independently distributed observations $a_i$, the maximum log-likelihood of a corresponding Gaussian regression model can be written as \cite{Konishi1995}:
\begin{equation}
\mathfrak{L}_{\rm normal} = - \frac{n}{2} \left[ \ln\!\left(2\pi \hat \sigma^2\right) + 1 \right] ,
\end{equation}
with
\begin{equation}
\hat \sigma^2 = \frac{1}{n} \sum_{i=1}^n {\left(\frac{a_i - b_i}{\rm eV}\right)}^2 \label{eq:sigma2}.
\end{equation} 
$b_i$ are the model predictions. Choosing a different energy unit than eV for normalization in Eq.~\ref{eq:sigma2} merely leads to a constant offset of the AIC, not affecting the relative ranking of the models. With decreasing $\hat \sigma^2$, {\it i.e.}\/, with increasing goodness of fit, the term $-2\mathfrak{L}_{\rm normal}$ in the AIC (\ref{eq:aic}) decreases. The term $2\Phi$, on the other hand, is an overfit penalty for the number of parameters of the model. The selection strategy is thus to prefer models with small AIC.

The observations are experimental oxide formation energies and the computed values from Eq.~\ref{eq:eform} the model predictions, both normalized for each oxide as described in the following. The more atoms there are per formula unit, the larger typically the error in the predicted formation energy of the compound. Cohesive energies of solids that are well described within the GGA, {\it e.g.}\/, have typical errors of 0.1~eV--0.2~eV per atom \cite{BrownJCC2021,ZhangNJP2018}. The considered oxides have a relatively wide range of ratios $\kappa$ of number of atoms $N_{\rm A}$ per formula unit to Hubbard-corrected sites $N_{\rm H}$ per formula unit ranging from 2 to 8, and oxides with larger $\kappa$ typically will have a larger GGA contribution to the error in the predicted formation energy per formula unit. If $a_i$ and $b_i$ were normalized by dividing by $N_{\rm H}$ or by number of formula units, residual errors for oxides with larger ratio $\kappa$ or larger formula units would be weighted more strongly in Eq.~\ref{eq:sigma2}, respectively. The $f^z_{\rm GP}$ could potentially be trained to compensate for relatively large GGA contributions to the errors, which is undesired. If $a_i$ and $b_i$ instead included a factor $1/N_{\rm A}$, the $E_{\rm off}(U,\rho)$ would have a small relative weight for oxides with large ratio $\kappa$. As a compromise, the experimental and computed oxide formation energies per formula unit are divided by $\sqrt{N_{\rm A}}$, ensuring that all $E_{\rm off}(U,\rho)$ have appreciable, balanced relative weight in the residuals $a_i-b_i$. With these normalized residuals $a_i-b_i$, the correspondingly normalized mean-absolute error (MAE) is defined: 
\begin{equation}
\Delta \tilde E_{\rm MAE} = \frac{1}{n} \sum_{i=1}^n \left|a_i - b_i\right| \label{eq:mae} ,
\end{equation}
which will be presented in addition to the model AIC in Sec.~\ref{sec:results}.

\begin{table}
\caption{Complexity weights $w_j$ for variables, numerical coefficients, and mathematical operations in the programs. Divisions have higher weight than additions and multiplications, to favor mathematical simplicity of the programs. The variables are ${\rm var}_1=\sum_\sigma {\rm Tr} \rho^\sigma$ and ${\rm var}_2=\sum_\sigma {\rm Tr} (\rho^\sigma)^2$, both with a weight of one. No weight is assigned for taking the square root of these variables. Other powers of the variables can only be expressed as products (or divisions) with corresponding penalty weights for multiplications and variable occurrences.\label{tab:weights}}
\centering
\footnotesize
\begin{tabular}{@{}lcccccc}
\br
& variable & integer & non-integer coeff. & add, subtract & multiply & divide\\
\mr
$j$ & 1 & 2 & 3 & 4 & 5 & 6\\
$w_j$ & 1 & 0 & 2 & 1 & 1 & 2\\
\br
\end{tabular}
\end{table}

The complexity term $2\Phi$ in Eq.~\ref{eq:aic} is generalized to not only account for and penalize the number of numerical parameters, but also mathematical operations:
\begin{equation}
\Phi = \sum_j w_j \varphi_j \label{eq:complexity}.
\end{equation}
$w_j$ is a weight for a mathematical operation, variable, or numerical coefficient (see Tab.~\ref{tab:weights}) and $\varphi_j$ the corresponding occurrence in the considered model (or program) $f^z_{\rm GP}$.

Using the resulting AIC as a fitness function evaluated on a random selection for each of the programs of half of the 66 TM oxide reference data, $k$-tournament selection \cite{Affenzeller2009} is employed as implemented in the gplearn code \cite{StephensGP2019}. In this procedure, $k$ programs out of the current generation (here $k$=25 out of a total population size of 100 per generation) are randomly selected, from which the program with lowest AIC is selected for crossover and mutation operations (operations named in analogy to biological genetic evolution). With 70\% probability, the winner from a second $k$-tournament is used for crossover of terms between the two programs, or with 25\% probability, random mutations are performed as described in Ref.~\cite{StephensGP2019}, and the resulting new program becomes part of the next generation. With 5\% probability, the tournament winner is promoted to the next generation unchanged. The initial generation consists of 30 copies of program (\ref{eq:start}) and 70 randomly generated $f^z_{\rm GP}$, and the programs are evolved for $10^5$ generations.

Beginning from the last generation, $10^5$ further generations are evolved, where now all programs in each generation are optimized numerically (using the Nelder-Mead simplex algorithm \cite{NelderMeadCJ1965,GaoHanCOA2012} as implemented in scipy \cite{scipy}) and simplified symbolically (using the sympy code \cite{sympy}) in addition to the genetic operations. Fractions in programs are normalized to equivalent fractions so that the coefficient of the lowest power in the variables in the denominator is one. To constrain the search space for the final $10^5$ generations, a numerically large penalty is added to the AIC for models that do not vanish for integer density matrix eigenvalues, {\it i.e.}\/, where $E_U$ vanishes, the DFT and DFT+$U$ energy functionals coincide, and thus no correction term for total energy comparison is required. A modified version of the gplearn code with unique weights for mathematical operations, coefficients, and variables and integrated symbolic and numerical optimization can be found on-line \cite{gplearnmod}.

\section{Results and discussion\label{sec:results}}

\begin{figure}
\includegraphics[width=\textwidth,keepaspectratio=true]{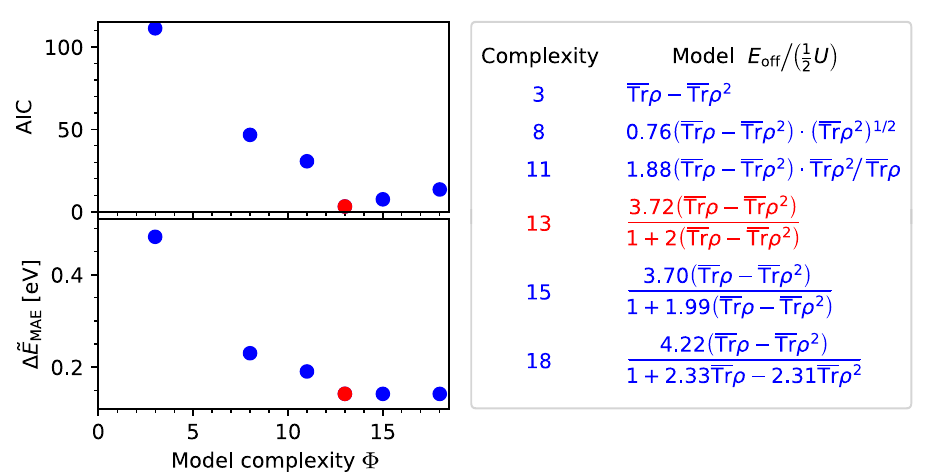}
\caption{
Results of the genetic programming model search. The AIC (Eq.~\ref{eq:aic}) and the residual error $\Delta\tilde{E}_{\rm MAE}$ (Eq.~\ref{eq:mae}) {\it vs.}~model complexity $\Phi$ (Eq.~\ref{eq:complexity}) are shown. In the equations describing the models, $\overline{\rm Tr}$ is short for the trace over the $d$-orbital degrees of freedom and the sum over the collinear spin degree of freedom $\sigma$. Only the models with lowest AIC for a given complexity and with an AIC$<$112 are shown. The model with lowest AIC is highlighted with red color.\label{fig:modelsearch}}
\end{figure}

From the constrained genetic programming search for models vanishing for integer density matrix eigenvalues, the programs with lowest AIC at a given complexity from the $10^5$ generations are shown in Fig.~\ref{fig:modelsearch}. All models found satisfying this constraint contain the factor $\sum_\sigma {\rm Tr}\rho^\sigma-{\rm Tr}(\rho^\sigma)^2$, which vanishes for integer eigenvalues of $\rho^\sigma$.

The simplest program is to subtract $E_U$ from the GGA+$U$ total energy, yielding a relatively high $\Delta\tilde{E}_{\rm MAE}$ of about 0.5~eV. Multiplying this model with a pre-factor increases the complexity $\Phi$ to 6. The resulting model $1.17\cdot(\sum_\sigma {\rm Tr}\rho^\sigma-{\rm Tr}(\rho^\sigma)^2)$ lowers $\Delta\tilde{E}_{\rm MAE}$ by only 6~meV, while the AIC increases by almost 6, thus identifying this model as an overfit. Significant improvements of the AIC and $\Delta\tilde{E}_{\rm MAE}$ are achieved by programs further multiplying $E_{\rm off}$ by $\sqrt{\sum_\sigma {\rm Tr}(\rho^\sigma)^2}$ or $\sum_\sigma {\rm Tr}(\rho^\sigma)^2\,/\,\sum_\sigma {\rm Tr}\rho^\sigma$, respectively.

The program with lowest AIC and hence optimal goodness of fit {\it vs.}\ complexity trade-off yields the correction term
\begin{equation}
E^{\rm opt}_{\rm off}(U_I,\rho_I) = 1.86\,U_I \frac{\sum_\sigma {\rm Tr}\rho_I^\sigma - {\rm Tr}{\left(\rho_I^\sigma\right)}^2}{1 + 2\left( \sum_\sigma {\rm Tr}\rho_I^\sigma - {\rm Tr}{\left(\rho_I^\sigma\right)}^2\right)} \label{eq:winningmodel},
\end{equation}
where the site index $I$ is not suppressed to emphasize that this model with site-specific $U_I$ Hubbard parameter strength is to be subtracted from GGA+$U$ total energies for every Hubbard-corrected site $I$ in the system.

The next more complex model in Fig.~\ref{fig:modelsearch} has a non-integer and numerically optimized coefficient in the denominator (leading to a complexity increase of 2), but the optimized coefficient is almost an integer, and the model is approximately equal to (\ref{eq:winningmodel}), and the AIC increases. The most complex model in Fig.~\ref{fig:modelsearch} further has slightly different numerical coefficients for $\sum_\sigma {\rm Tr}\rho^\sigma$ and $(\sum_\sigma {\rm Tr}\rho^\sigma)^2$ in the denominator, which increases the AIC, identifying a tendency towards overfitting. Eq.~\ref{eq:winningmodel} is thus the model identified as optimal for computing offsets to GGA+$U$ total energies for comparison at different $U$-values, including ($U=0$) GGA total energies.

\begin{figure}
\includegraphics[width=\textwidth,keepaspectratio=true]{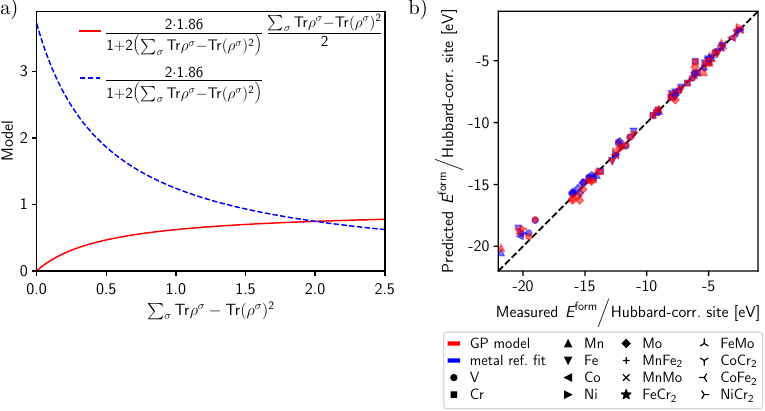}
\caption{Functional form of the model $E^{\rm opt}_{\rm off}$ (\ref{eq:winningmodel}) with lowest AIC found (a): The solid red line shows the model (in units of the site-dependent Hubbard parameter) in dependence of the physically possible range of $\sum_\sigma {\rm Tr}\rho^\sigma - {\rm Tr}(\rho^\sigma)^2$ for $d$-subshell occupations. The dashed blue line shows the ratio $E^{\rm opt}_{\rm off}/E_U$. For $\sum_\sigma {\rm Tr}\rho^\sigma - {\rm Tr}(\rho^\sigma)^2=1.36$ this scaling function has a value of one where $E^{\rm opt}_{\rm off}$ exactly cancels $E_U$. Performance of the model for predicting heat of formation for the 66 benchmark TM oxides (b): parity plot of the predicted oxide formation energies divided by the number of Hubbard-corrected sites {\it vs.}\ experimental references (red symbols). For comparison, the predictions from GGA+$U$ simulations with average, fitted $U$-parameters and fitted metal references is shown (blue symbols). The symbol shapes indicate the ionic species of the Hubbard-corrected sites.\label{fig:model}}
\end{figure}

$E^{\rm opt}_{\rm off}$ only depends on the density matrices through the difference $\Delta = \sum_\sigma {\rm Tr}\rho^\sigma-{\rm Tr}(\rho^\sigma)^2$, and this dependence is depicted in Fig.~\ref{fig:model}a. The model can be considered a rescaling of $E_U$. For differences $0<\Delta<1.36$, $E^{\rm opt}_{\rm off}$ is increased over $E_U$, while for larger differences, it is decreased relative to $E_U$.

In Fig.~\ref{fig:model}b, the performance of the model is compared to GGA+$U$ calculations with average, fitted $U$-values. The average $U$-values are taken from the Materials Project \cite{JainAPLM2013} and are listed in Tab.~S3. Constant energy offsets for each TM species are fitted to the binary $3d$ and Mo oxide heats of formation, following the procedure in Ref.~\cite{JainPRB2011}. With the seven different TM ions considered and with the empirical, fitted $U$-values and energy offsets, there are thus effectively 14 fitting parameters. $E^{\rm opt}_{\rm off}$ with the first-principles, site-dependent $U$-values and thus significantly fewer degrees of freedom (through its functional form and numerical coefficient) leads to similar residual errors in comparison to the experimental benchmarks, showing generally good agreement (experimental and computed TM oxide heats of formation listed in Tab.~S2). With the similar quality of prediction for these bulk oxide formation benchmark cases, the method presented here holds the promise of enabling predictions of TM oxide formation energies and other reactions energies between transition metals and their oxides where experimental benchmark data for fitting is lacking.

The heats of formation of the ruthenates Ca$_2$RuO$_4$ and Y$_2$Ru$_2$O$_7$ serve here as well-defined benchmark cases; other Ru oxides, such as RuO$_2$, CaRuO$_3$, SrRuO$_3$, and Sr$_2$RuO$_4$ are not considered here, as these oxides are metallic \cite{LongoJAP1968} (Sr$_2$RuO$_4$ is metallic at temperatures below $\sim$130~K and superconducting below $\sim$1~K \cite{MaenoNat1994}) and should thus not be treated at the GGA$+U$ level of theory. As there are only two Hubbard-corrected Ru oxides considered here, no energy offset nor average $U$-value is fitted for Ru, but only the first-principles $U$-value approach based on $E^{\rm opt}_{\rm off}$ is applied to compute the Ru oxide heats of formation. Calculation of the experimental benchmark values is explained in Sec.~\ref{sec:ruoxbench}.

Ca$_2$RuO$_4$ is an insulator at ambient conditions \cite{NakatsujiJPSJ1997}. With powder diffraction-based structural refinements of the $Pbca$-phase at 295~K from Ref.~\cite{BradenPRB1998} as a starting guess, the structure of Ca$_2$RuO$_4$ is optimized iteratively at the GGA+$U$ level with self-consistent $U$-parameters as described in Sec.~\ref{sec:upractice}. Antiferromagnetic order with a propagation vector of $(1,0,0)$ (corresponding to the $A$-centered mode \cite{BradenPRB1998}) is considered (the energy of the $B$-centered mode with propagation vector $(0,1,0)$ is found to be within {$\sim$}$10^{-4}$~eV/atom and is thus energetically almost indistinguishable from the $A$-centered phase within the accuracy of the simulations, allowing the lattice to relax for both the $A$- and $B$-centered modes). A Hubbard-parameter strength $U^{\rm Ru}_I${$\approx$4.5~eV} is calculated with Eq.~\ref{eq:uresp}. With the corresponding GGA+$U$ total energy of Ca$_2$RuO$_4$ and GGA total energies for the reactants and applying corrections (\ref{eq:eo2}) and (\ref{eq:winningmodel}), a heat of formation of \mbox{$-2.29$}~eV/atom is predicted, which is in good agreement with the experimental result of \mbox{$-2.35$}~eV/atom.

\begin{figure}
\centering
\includegraphics[width=2.60in,keepaspectratio=true]{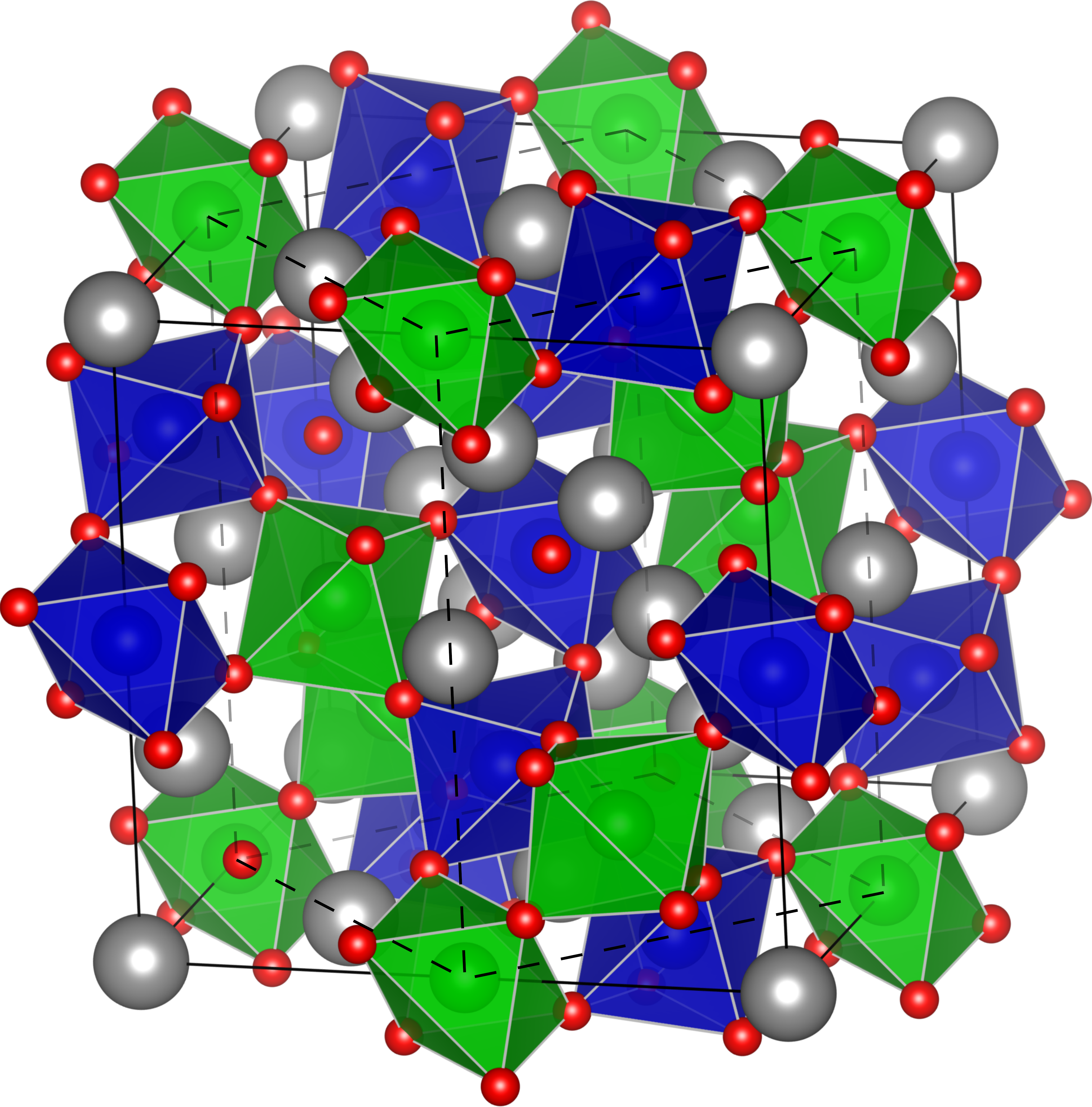}
\caption{VESTA \cite{MommaJAC2011} rendition of the pyrochlore crystal structure of Y$_2$Ru$_2$O$_7$. Gray spheres represent Y and red spheres O. The octahedra show the coordination of the Hubbard-corrected Ru-sites, with green and blue color for the two different magnetic moment orientations used in the collinear-spin GGA+$U$ simulation, respectively. The dashed lines indicate the primitive magnetic unit cell.\label{fig:y2ru2o7}}
\end{figure}

Y$_2$Ru$_2$O$_7$ has pyrochlore structure (spacegroup $Fd\bar 3m$ \cite{KannoJSSC1993}) and is an insulator with magnetic frustration and Curie-Weiss-like behavior above $\sim$100~K \cite{YoshiiJPSJ1999}. Below $\sim$76~K, Y$_2$Ru$_2$O$_7$ shows antiferromagnetic, non-collinear spin ordering \cite{BlundellPRB2008}. Here, the magnetic structure of Y$_2$Ru$_2$O$_7$ is approximated with collinear spins, and different spin orderings in a supercell containing eight formula units as depicted in Fig.~\ref{fig:y2ru2o7} were compared energetically at the GGA+$U$ level. The low-energy antiferromagnetic phase used in the following is depicted in Fig.~\ref{fig:y2ru2o7} (dashed lines indicate the primitive magnetic cell containing four formula units). After iterative structural optimization and computation of the first-principles Hubbard-parameter strength, a parameter $U^{\rm Ru}_I${$\approx$4.3~eV} is found. The computed heat of formation is \mbox{$-2.36$}~eV/atom, comparing well to the experimental estimate of \mbox{$-2.44$}~eV/atom. This \mbox{$\lesssim$0.1}~eV/atom error is within the range of errors found in Ref.~\cite{WangPRB2006} for the heat of formation of oxides not containing transition metal ions (and thus not requiring Hubbard corrections) with a fitted O$_2$ gas reference, and within the range of GGA reaction energy errors involving metallic TM phases in general \cite{ZhangNJP2018,BrownJCC2021,SharadaPRB2019}. While oxidation state-specific TM references fitted to experiments can reduce per atom oxide reaction energy errors significantly \cite{AykolPRB2014}, having established here an approach to correcting mixed GGA and GGA+$U$ total energy differences with site-specific, first-principles Hubbard parameters, which leads to errors typical for GGA-only estimates for reaction energies without localization errors, will enable oxide reaction energy predictions in the absence of experimental reference data.

\section{Conclusion}

The presented approach enables the prediction of transition metal oxide formation energies and other reaction energies involving oxide and metallic phases with Hubbard corrections computed from first principles. Comparison of total energies from Hamiltonians with different Hubbard parameter strengths including $U=0$ is facilitated with a site-projected density matrix-dependent model, that was found through genetic programming against experimental transition metal oxide formation energies. This mathematically simple model does not involve fitting parameters to be adjusted to transition metal ion types or oxidation states. Unlike previous Hubbard-corrected approaches to TM oxide heat of formation prediction requiring experimental benchmark data to fit energetic corrections and average Hubbard strengths, the method presented here can thus also be used to compute transition metal oxide reaction energies when such experimental fitting targets are lacking, which will be of particular importance for oxide interfaces and defect chemistry.

\section{Acknowledgment}

This research was supported by the U.S.\ Department of Energy, Office of Science, Office of Basic Energy Sciences, Chemical Sciences, Geosciences, and Biosciences Division, Catalysis Science Program to the SUNCAT Center for Interface Science and Catalysis.

\bibliography{main}

\end{document}


{\Huge \textsf{\vspace{-1.82ex}\\ {\Large Supplementary Information}\\ ~\vspace{0.75ex}\\ Hubbard-corrected oxide formation enthalpies\\ without adjustable parameters\\
~\vspace{0.75ex}\\ {\Large J Voss}}}~\hfill{}January 26, 2022\vspace{1.5ex}\\
SUNCAT Center for Interface Science and Catalysis\\ SLAC National Accelerator Laboratory, Menlo Park, CA 94025, USA\\
E-mail: \texttt{vossj@slac.stanford.edu}\\
~\vspace{2ex}\\
{\Large \textsf{1. Computed transition metal oxide \emph{U}-parameters and magnetic ordering}}
~\vspace{2ex}\\
\begin{tabular}{|l|c|c|l|}
\hline
Oxide~~~~~~\, & Mag. & \,Spg.\, & Transition metal sites with $U$\hspace{-0.1em}-parameters [eV]~\hspace{96.75096pt}~\\
\hline
BaV$_2$O$_6$ & NM & 21 & \tmsite{V,}{$4k$:}{3.982~~\,} \tmsite{V,}{$8l$:}{3.935~~\,} \\
Ca$_2$V$_2$O$_7$ & NM & 2 & \tmsite{V,}{$2i$:}{4.771~~\,} \tmsite{V,}{$2i$:}{4.544~~\,} \\
Ca$_3$V$_2$O$_8$ & NM & 2 & \tmsite{V,}{$2i$:}{4.332~~\,} \\
CaV$_2$O$_6$ & NM & 12 & \tmsite{V,}{$4i$:}{4.746~~\,} \\
Mg$_2$V$_2$O$_7$ & NM & 2 & \tmsite{V,}{$2i$:}{4.637~~\,} \tmsite{V,}{$2i$:}{4.434~~\,} \\
MgV$_2$O$_6$ & NM & 8 & \tmsite{V,}{$2a$:}{4.788~~\,} \tmsite{V,}{$2a$:}{4.700~~\,} \\
Na$_3$VO$_4$ & NM & 31 & \tmsite{V,}{$2a$:}{4.374~~\,} \\
Na$_4$V$_2$O$_7$ & NM & 15 & \tmsite{V,}{$8f$:}{4.362~~\,} \tmsite{V,}{$8f$:}{4.361~~\,} \tmsite{V,}{$8f$:}{4.348~~\,} \tmsite{V,}{$8f$:}{4.323} \\
NaVO$_3$ & NM & 15 & \tmsite{V,}{$8f$:}{4.573~~\,} \\
V$_2$O$_3$$^*$ & AFM & 15 & \tmsite{V,}{$8f$:}{6.027~~\,} \\
V$_2$O$_5$ & NM & 59 & \tmsite{V,}{$4e$:}{4.819~~\,} \\
VO & AFM & 166 & \tmsite{V,}{$3a$:}{5.836~~\,} \\
VO$_2$$^*$ & AFM & 136 & \tmsite{V,}{$2b$:}{5.387~~\,} \\
CaCr$_2$O$_4$ & FM & 62 & \tmsite{Cr,}{$4c$:}{5.150~~\,} \tmsite{Cr,}{$4c$:}{5.105~~\,} \\
Cr$_2$O$_3$ & FM & 167 & \tmsite{Cr,}{$12c$:}{5.424~~\,} \\
CrO$_3$ & FM & 40 & \tmsite{Cr,}{$4b$:}{5.644~~\,} \\
Cs$_2$CrO$_4$ & NM & 62 & \tmsite{Cr,}{$4c$:}{4.306~~\,} \\
CuCr$_2$O$_4$$^*$ & FM & 227 & \tmsite{Cr,}{$16c$:}{7.156~~\,} \\
K$_2$CrO$_4$ & NM & 62 & \tmsite{Cr,}{$4c$:}{4.294~~\,} \\
MgCr$_2$O$_4$ & FM & 227 & \tmsite{Cr,}{$16d$:}{4.899~~\,} \\
Na$_2$CrO$_4$ & NM & 63 & \tmsite{Cr,}{$4c$:}{4.611~~\,} \\
NaCrO$_2$ & FM & 166 & \tmsite{Cr,}{$3a$:}{4.612~~\,} \\
ZnCr$_2$O$_4$ & FM & 227 & \tmsite{Cr,}{$16c$:}{4.938~~\,} \\
Mn$_2$O$_3$ & FM & 61 & \tmsite{Mn,}{$4a$:}{6.025~~\,} \tmsite{Mn,}{$4b$:}{5.130~~\,} \tmsite{Mn,}{$8c$:}{5.362~~\,} \tmsite{Mn,}{$8c$:}{5.346} \\
& & & \tmsite{Mn,}{$8c$:}{5.258~~\,} \\
Mn$_2$TiO$_4$ & FM & 95 & \tmsite{Mn,}{$4b$:}{3.769~~\,} \tmsite{Mn,}{$4c$:}{3.591~~\,} \\
Mn$_3$O$_4$ & FM & 141 & \tmsite{Mn,}{$4b$:}{4.347~~\,} \tmsite{Mn,}{$8c$:}{4.624~~\,} \\
MnAl$_2$O$_4$ & FM & 227 & \tmsite{Mn,}{$8a$:}{2.845~~\,} \\
MnO & AFM & 166 & \tmsite{Mn,}{$3a$:}{3.382~~\,} \\
MnO$_2$ & AFM & 87 & \tmsite{Mn,}{$4h$:}{4.786~~\,} \\
MnTiO$_3$ & FM & 148 & \tmsite{Mn,}{$6c$:}{3.789~~\,} \\
\hline
\end{tabular}\medskip\\
Table S1: \emph{Continued on next page.}\pagebreak\\
\begin{tabular}{|l|c|c|l|}
\hline
Oxide~~~~~~\, & Mag. & \,Spg.\, & Transition metal sites with $U$\hspace{-0.1em}-parameters [eV]~\hspace{96.75096pt}~\\
\hline
Ca$_2$Fe$_2$O$_5$$^*$ & FM & 62 & \tmsite{Fe,}{$4b$:}{7.640~~\,} \tmsite{Fe,}{$4c$:}{7.036~~\,} \\
CaFe$_2$O$_4$$^*$ & FM & 62 & \tmsite{Fe,}{$4c$:}{7.534~~\,} \tmsite{Fe,}{$4c$:}{7.486~~\,} \\
CuFe$_2$O$_4$$^*$ & FM & 227 & \tmsite{Fe,}{$16d$:}{8.421~~\,} \\
CuFeO$_2$$^*$ & FM & 166 & \tmsite{Fe,}{$3b$:}{7.303~~\,} \\
Fe$_2$O$_3$ & AFM & 167 & \tmsite{Fe,}{$12c$:}{7.921~~\,} \\
Fe$_2$TiO$_4$$^*$ & FM & 74 & \tmsite{Fe,}{$4d$:}{7.366~~\,} \tmsite{Fe,}{$4e$:}{7.246~~\,} \\
Fe$_3$O$_4$ & FiM & 166 & \tmsite{Fe,}{$3b$:}{8.289~~\,} \tmsite{Fe,}{$6c$:}{7.836~~\,} \tmsite{Fe,}{$9e$:}{8.316~~\,} \\
FeAl$_2$O$_4$$^*$ & FM & 227 & \tmsite{Fe,}{$8b$:}{8.917~~\,} \\
FeO$^*$ & AFM & 139 & \tmsite{Fe,}{$2a$:}{6.373~~\,} \\
FeTiO$_3$$^*$ & FM & 148 & \tmsite{Fe,}{$6c$:}{7.683~~\,} \\
LiFeO$_2$$^*$ & FM & 141 & \tmsite{Fe,}{$4a$:}{7.373~~\,} \\
NaFeO$_2$ & FM & 166 & \tmsite{Fe,}{$3b$:}{7.233~~\,} \\
ZnFe$_2$O$_4$$^*$ & FM & 227 & \tmsite{Fe,}{$16d$:}{7.537~~\,} \\
Co$_3$O$_4$ & AFM & 227 & \tmsite{Co,}{$8a$:}{6.718~~\,} \tmsite{Co,}{$16d$:}{6.558~~\,} \\
CoAl$_2$O$_4$ & FM & 227 & \tmsite{Co,}{$8b$:}{4.637~~\,} \\
CoO & AFM & 160 & \tmsite{Co,}{$3a$:}{6.842~~\,} \\
CoTiO$_3$ & FM & 148 & \tmsite{Co,}{$6c$:}{6.997~~\,} \\
NiAl$_2$O$_4$ & FM & 74 & \tmsite{Ni,}{$4c$:}{3.171~~\,} \\
NiO & AFM & 225 & \tmsite{Ni,}{$4b$:}{3.293~~\,} \\
NiTiO$_3$ & FM & 148 & \tmsite{Ni,}{$6c$:}{3.578~~\,} \\
BaMoO$_4$ & NM & 88 & \tmsite{Mo,}{$4b$:}{2.282~~\,} \\
CaMoO$_4$ & NM & 88 & \tmsite{Mo,}{$4b$:}{2.479~~\,} \\
Cs$_2$MoO$_4$ & NM & 62 & \tmsite{Mo,}{$4c$:}{2.222~~\,} \\
MgMoO$_4$ & NM & 12 & \tmsite{Mo,}{$4g$:}{2.546~~\,} \tmsite{Mo,}{$4i$:}{2.532~~\,} \\
MoO$_2$$^*$ & AFM & 136 & \tmsite{Mo,}{$2a$:}{3.340~~\,} \\
MoO$_3$ & NM & 14 & \tmsite{Mo,}{$4e$:}{2.604~~\,} \tmsite{Mo,}{$4e$:}{2.603~~\,} \\
Na$_2$Mo$_2$O$_7$ & NM & 64 & \tmsite{Mo,}{$8e$:}{2.833~~\,} \tmsite{Mo,}{$8f$:}{2.515~~\,} \\
Na$_2$MoO$_4$ & NM & 227 & \tmsite{Mo,}{$8a$:}{2.448~~\,} \\
SrMoO$_4$ & NM & 88 & \tmsite{Mo,}{$4b$:}{2.457~~\,} \\
FeCr$_2$O$_4$$^*$ & FM & 227 & \tmsite{Fe,}{$8a$:}{8.873~~\,} \tmsite{Cr,}{$16d$:}{5.384~~\,} \\
CoCr$_2$O$_4$ & FM & 227 & \tmsite{Co,}{$8b$:}{5.468~~\,} \tmsite{Cr,}{$16c$:}{5.020~~\,} \\
NiCr$_2$O$_4$$^*$ & FM & 141 & \tmsite{Ni,}{$4b$:}{3.552~~\,} \tmsite{Cr,}{$8c$:}{5.444~~\,} \\
MnFe$_2$O$_4$$^*$ & FM & 74 & \tmsite{Fe,}{$4a$:}{6.379~~\,} \tmsite{Mn,}{$4d$:}{2.857~~\,} \tmsite{Fe,}{$4e$:}{7.578~~\,} \\
MnMoO$_4$ & FM & 12 & \tmsite{Mo,}{$4g$:}{2.671~~\,} \tmsite{Mn,}{$4h$:}{3.881~~\,} \tmsite{Mn,}{$4i$:}{3.972~~\,} \tmsite{Mo,}{$4i$:}{2.636} \\
CoFe$_2$O$_4$$^*$ & FM & 227 & \tmsite{Co,}{$8a$:}{6.099~~\,} \tmsite{Fe,}{$16d$:}{7.654~~\,} \\
FeMoO$_4$$^*$ & FM & 12 & \tmsite{Mo,}{$4g$:}{2.808~~\,} \tmsite{Fe,}{$4h$:}{6.808~~\,} \tmsite{Fe,}{$4i$:}{8.471~~\,} \tmsite{Mo,}{$4i$:}{2.545} \\
Ca$_2$RuO$_4$$^*$ & AFM & 61 & \tmsite{Ru,}{$4a$:}{4.472~~\,} \\
Y$_2$Ru$_2$O$_7$$^*$ &$\!\!$AFM$^\dagger\!\!$& 227 & \tmsite{Ru,}{$16d$:}{4.286~~\,} \\
\hline
\end{tabular}\medskip\\
Table S1: \emph{Continued from previous page.} Transition metal site-dependent $U$-parameters and GGA+$U$ magnetic ordering (Mag.): ferromagnetic (FM), antiferromagnetic (AFM), ferrimagnetic (FiM), or (treated as) non-magnetic (NM). Crystallographic space group numbers (Spg.)\ according to the International Tables for Crystallography [1]. $^*\,$GGA (PBE [2]) electronic structure metallic or with vanishing bandgap: $U$-parameters computed from linear extrapolation to \mbox{GGA+$U\!=\!0$} from gapped GGA+$U$ electronic structures. All other $U$-parameters obtained from GGA calculations. $^\dagger\,$Frustrated magnetism in Y$_2$Ru$_2$O$_7$ approximated with supercell with net zero magnetic moment. See main article for details. \vfill{}
\clearpage

{\Large \textsf{\vspace{-4ex}\\ {2. Transition metal oxide formation energies}}}
~\vspace{2ex}\\
\begin{tabular}{|l|c|c|c|}
\hline
Oxide~~~~~~\, & ${\Delta}E_{\rm GP}$ & $\!\!{\Delta}E_{\rm metref}\!\!$ & ${\Delta}E_{\rm exp}$ \\
 & ~~~[eV]~~~ & ~~~[eV]~~~ & ~~~[eV]~~~ \\
\hline
BaV$_2$O$_6$ & -23.637 & -23.795 & -23.361 \\
Ca$_2$V$_2$O$_7$ & -31.463 & -31.311 & -31.896 \\
Ca$_3$V$_2$O$_8$ & -38.401 & -37.846 & -39.079 \\
CaV$_2$O$_6$ & -23.812 & -23.867 & -24.074 \\
Mg$_2$V$_2$O$_7$ & -28.947 & -28.789 & -29.316 \\
MgV$_2$O$_6$ & -22.402 & -22.309 & -22.726 \\
Na$_3$VO$_4$ & -17.842 & -17.914 & -19.020 \\
Na$_4$V$_2$O$_7$ & -29.829 & -29.575 & -30.216 \\
NaVO$_3$ & -11.559 & -11.690 & -12.296 \\
V$_2$O$_3$ & -12.793 & -12.811 & -12.643 \\
V$_2$O$_5$ & -15.761 & -15.849 & -16.011 \\
VO & \hspace{0.5em}-4.275 & \hspace{0.5em}-4.370 & \hspace{0.5em}-4.483 \\
VO$_2$ & \hspace{0.5em}-7.417 & \hspace{0.5em}-7.421 & \hspace{0.5em}-7.413 \\
CaCr$_2$O$_4$ & -18.892 & -18.822 & -19.001 \\
Cr$_2$O$_3$ & -11.915 & -12.012 & -11.810 \\
CrO$_3$ & \hspace{0.5em}-5.094 & \hspace{0.5em}-5.040 & \hspace{0.5em}-6.076 \\
Cs$_2$CrO$_4$ & -14.711 & -14.537 & -14.674 \\
CuCr$_2$O$_4$ & -13.629 & -13.570 & -13.431 \\
K$_2$CrO$_4$ & -14.617 & -14.438 & -14.309 \\
MgCr$_2$O$_4$ & -18.364 & -18.382 & -18.438 \\
Na$_2$CrO$_4$ & -13.981 & -13.916 & -13.746 \\
NaCrO$_2$ & \hspace{0.5em}-8.954 & \hspace{0.5em}-9.021 & \hspace{0.5em}-9.100 \\
ZnCr$_2$O$_4$ & -15.947 & -15.874 & -16.099 \\
Mn$_2$O$_3$ & -10.317 & -10.063 & \hspace{0.5em}-9.912 \\
Mn$_2$TiO$_4$ & -17.782 & -18.117 & -18.068 \\
Mn$_3$O$_4$ & -14.302 & -14.256 & -14.329 \\
MnAl$_2$O$_4$ & -20.128 & -20.507 & -21.782 \\
MnO & \hspace{0.5em}-3.542 & \hspace{0.5em}-3.898 & \hspace{0.5em}-3.968 \\
MnO$_2$ & \hspace{0.5em}-5.936 & \hspace{0.5em}-5.461 & \hspace{0.5em}-5.377 \\
MnTiO$_3$ & -13.940 & -14.245 & -14.051 \\
Ca$_2$Fe$_2$O$_5$ & -21.974 & -21.439 & -22.120 \\
CaFe$_2$O$_4$ & -15.382 & -15.108 & -15.348 \\
CuFe$_2$O$_4$ & -10.167 & \hspace{0.5em}-9.759 & -10.032 \\
CuFeO$_2$ & \hspace{0.5em}-5.348 & \hspace{0.5em}-5.204 & \hspace{0.5em}-5.285 \\
\hline
\end{tabular}~\hfill{}~%
\begin{tabular}{|l|c|c|c|}
\hline
Oxide~~~~~~\, & ${\Delta}E_{\rm GP}$ & $\!\!{\Delta}E_{\rm metref}\!\!$ & ${\Delta}E_{\rm exp}$ \\
 & ~~~[eV]~~~ & ~~~[eV]~~~ & ~~~[eV]~~~ \\
\hline
Fe$_2$O$_3$ & \hspace{0.5em}-8.887 & \hspace{0.5em}-8.751 & \hspace{0.5em}-8.563 \\
Fe$_2$TiO$_4$ & -15.720 & -15.880 & -15.476 \\
Fe$_3$O$_4$ & -11.806 & -11.442 & -11.603 \\
FeAl$_2$O$_4$ & -18.819 & -18.556 & -20.378 \\
FeO & \hspace{0.5em}-2.719 & \hspace{0.5em}-2.829 & \hspace{0.5em}-2.715 \\
FeTiO$_3$ & -13.141 & -13.081 & -12.789 \\
LiFeO$_2$ & \hspace{0.5em}-7.831 & \hspace{0.5em}-7.697 & \hspace{0.5em}-7.973 \\
NaFeO$_2$ & \hspace{0.5em}-7.303 & \hspace{0.5em}-7.167 & \hspace{0.5em}-7.219 \\
ZnFe$_2$O$_4$ & -12.270 & -12.068 & -12.163 \\
Co$_3$O$_4$ & \hspace{0.5em}-9.357 & \hspace{0.5em}-9.419 & \hspace{0.5em}-9.405 \\
CoAl$_2$O$_4$ & -18.997 & -19.185 & -20.253 \\
CoO & \hspace{0.5em}-2.510 & \hspace{0.5em}-2.453 & \hspace{0.5em}-2.484 \\
CoTiO$_3$ & -12.639 & -12.597 & -12.544 \\
NiAl$_2$O$_4$ & -18.596 & -18.959 & -19.979 \\
NiO & \hspace{0.5em}-2.280 & \hspace{0.5em}-2.496 & \hspace{0.5em}-2.496 \\
NiTiO$_3$ & -12.339 & -12.576 & -12.482 \\
BaMoO$_4$ & -16.289 & -15.700 & -15.428 \\
CaMoO$_4$ & -16.128 & -15.630 & -15.989 \\
Cs$_2$MoO$_4$ & -15.941 & -15.314 & -15.562 \\
MgMoO$_4$ & -14.752 & -14.266 & -14.461 \\
MoO$_2$ & \hspace{0.5em}-5.515 & \hspace{0.5em}-6.056 & \hspace{0.5em}-6.104 \\
MoO$_3$ & \hspace{0.5em}-8.166 & \hspace{0.5em}-7.779 & \hspace{0.5em}-7.694 \\
Na$_2$Mo$_2$O$_7$ & -24.103 & -23.062 & -24.359 \\
Na$_2$MoO$_4$ & -15.409 & -14.887 & -15.168 \\
SrMoO$_4$ & -16.292 & -15.780 & -16.009 \\
FeCr$_2$O$_4$ & -15.333 & -14.496 & -14.913 \\
CoCr$_2$O$_4$ & -14.843 & -14.974 & -14.917 \\
NiCr$_2$O$_4$ & -13.883 & -14.289 & -14.282 \\
MnFe$_2$O$_4$ & -12.366 & -12.007 & -12.707 \\
MnMoO$_4$ & -12.401 & -12.380 & -12.310 \\
CoFe$_2$O$_4$ & -11.153 & -11.094 & -11.295 \\
FeMoO$_4$ & -11.374 & -11.229 & -11.057 \\
Ca$_2$RuO$_4$ & -16.018 & --- & -16.458 \\
Y$_2$Ru$_2$O$_7$ & -25.954 & --- & -26.847 \\
\hline
\end{tabular}\medskip\\
Table S2: Transition metal oxide formation energies per formula unit computed using the presented genetic programming (GP) model with site-dependent $U$-parameters, using fitted metallic references (metref) and site-independent $U$-parameters, and as measured experimentally (exp). Ca$_2$RuO$_4$ experimental formation enthalpy from [3]; low-temperature and moderate temperature thermodynamic data from [4] and [5], respectively, used for extrapolation to 0~K. Y$_2$Ru$_2$O$_7$ formation enthalpy from [6] and low-temperature and moderate temperature thermodynamic data from [6-8]. Thermodynamic data for O$_2$ from [9]. Quantum chemistry corrections and zero-point energy of O$_2$ from [10]. Thermodynamic corrections for ZnFe$_2$O$_4$ from [11]. All other experimental thermodynamic data from [12]. Zero-point energies $E_{\rm ZP}$ subtracted from experimental energies. $E_{\rm ZP}$ for solids from Debye model fits. See main article for details.

{\Large \textsf{\vspace{-4ex}\\ {3. Average, site-independent \emph{U}-parameters from the Materials Project}}}
~\vspace{2ex}\\
\begin{tabular}{|l|c|}
\hline
Transition metal ion & $U$\hspace{-0.1em}-parameter [eV] \\
\hline
V & 3.25 \\
Cr & 3.7\hspace{0.5em} \\
Mn & 3.9\hspace{0.5em} \\
Fe & 5.3\hspace{0.5em} \\
Co & 3.32 \\
Ni & 6.2\hspace{0.5em} \\
Mo & 4.3\hspace{0.5em} \\
\hline
\end{tabular}\medskip\\
Table S3: Average fitted, site-independent (and thus also oxidation state-independent) $U$-parameters used in the Materials Project [13].

{\Large \textsf{\vspace{3ex}\\ {References}}}
~\vspace{-3ex}\\
\begin{enumerate}[label={[\arabic*]},leftmargin=*,itemsep=0.333333ex]
\item Hahn T, ed 2002 {\it International tables for crystallography}\/ Vol A, 5th ed (Dordrecht: Kluwer Academic Publishers)
\item Perdew P E, Burke K and Ernzerhof M 1996 \href{https://doi.org/10.1103/PhysRevLett.77.3865}{{\it Phys.\ Rev.\ Lett.}\/ {\bf 77} 3865}
\item Jacob K T, Lwin K T and Waseda Y 2003 \href{https://doi.org/10.1149/1.1557082}{{\it J.\ Electrochem.\ Soc.}\/ {\bf 150} E227}
\item Nakatsuji S 2000 \href{https://doi.org/10.11501/3179047}{\it Quasi-two-dimensional Mott transition system Ca$_{\mathit{2}-x}$Sr$_x$RuO$_\mathit{4}$} (Kyoto University: PhD Thesis)
\item Qi T F, Korneta O B, Parkin S, Hu J and Cao G 2012 \href{https://doi.org/10.1103/PhysRevB.85.165143}{{\it Phys.\ Rev.\ B}\/ {\bf 85} 165143}
\item Banerjee A 2019 \href{https://doi.org/10.1007/s10008-019-04268-8}{{\it J.\ Solid State Electrochem.}\/ {\bf 23} 1749}
\item Blacklock K, White H W and G{\"u}rmen E 1980 \href{https://doi.org/10.1063/1.440285}{{\it J. Chem. Phys.}\/ {\bf 73} 1966}
\item Taira N, Wakeshima M and Hinatsu Y 2000 \href{https://doi.org/10.1006/jssc.2000.8702}{{\it J.\ Solid State Chem.}\/ {\bf 152} 441}
\item Chase M W 1998 {\it NIST-JANAF thermochemical tables}\/ 4th ed (Woodbury, NY: American Institute of Physics)
\item Karton A, Sylvetsky N and Martin J M L 2017 \href{https://doi.org/10.1002/jcc.24854}{{\it J.\ Comput.\ Chem.}\/ {\bf 38} 2063}
\item Westrum E F and Grimes D 1957 \href{https://doi.org/10.1016/0022-3697(57)90046-X}{{\it J.\ Phys.\ Chem.\ Solids}\/ {\bf 3} 44}
\item Kubaschewski O, Spencer P J and Alcock C B 1993 {\it Materials thermochemistry}\/ 6th ed (Oxford: Pergamon Press)
\item Jain A, Ong S P, Hautier G, Chen W, Richards W D, Dacek S, Cholia S, Gunter D, Skinner D, Ceder G and Persson K A 2013 \href{https://doi.org/10.1063/1.4812323}{{\it APL Mater.}\/ {\bf 1} 011002}
\end{enumerate}